\documentclass{sf2a-conf2017}
\usepackage{graphicx}
\usepackage{hyperref}
\usepackage[]{natbib}
\usepackage{epstopdf}

\def\BibTeX{{\rm B\kern-.05em{\sc i\kern-.025em b}\kern-.08em
    T\kern-.1667em\lower.7ex\hbox{E}\kern-.125emX}}
\bibpunct{(}{)}{;}{a}{}{,}  %%%%%%%%%%%%%  A&A bibliography style
\begin{document}

\TitreGlobal{SF2A 2017}

%%-----------------------------------------------------------------
%%      the top matter
%%

\title{\emph{Gaia} DR1 completeness within 250\,pc and star formation history of the Solar neighbourhood}

\runningtitle{Star formation history of the Solar neighbourhood}

\author{Edouard J. Bernard}\address{Universit\'e C\^ote d’Azur, OCA, CNRS, Lagrange, France -- email: {\tt ebernard@oca.eu}}

%% Keep this line, even if the page will be settled afterwards.
\setcounter{page}{1}

%%-----------------------------------------------------------------

\maketitle

%%-----------------------------------------------------------------
%%        The abstract
%% 
%%  Warning!  within the abstract:
%%  - do not use macros.
%%  - do not use commands like: \cite, \citet, \citep ... etc.

\begin{abstract}
Taking advantage of the \emph{Gaia} DR1, we combined TGAS parallaxes with the \emph{Tycho-2} and APASS photometry to calculate the star formation history (SFH) of the solar neighbourhood within 250 pc using the colour-magnitude diagram fitting technique. Our dynamically-evolved SFH is in excellent agreement with that calculated from the \emph{Hipparcos} catalogue within 80 pc of the Sun, showing an enhanced star formation rate (SFR) in the past $\sim$4 Gyr. We then correct the SFR for the disc thickening with age to obtain a SFR that is representative of the whole solar cylinder, and show that even with an extreme correction our results are not consistent with an exponentially decreasing SFR as found by recent studies. Finally, we discuss how this technique can be applied out to $\sim$5 kpc thanks to the next \emph{Gaia} data releases, which will allow us to quantify the SFH of the thin disc, thick disc and halo \emph{in situ}.
\end{abstract}

%% Insert the keywords (to appear in the ADS indexing)
%% Keywords must be separated by a comma
\begin{keywords}
Hertzsprung-Russell diagram, Galaxy: disk, Galaxy: evolution, Galaxy: formation, solar neighbourhood
\end{keywords}

%%-----------------------------------------------------------------

\section{Introduction}
%%--------------------

Disc galaxies dominate the stellar mass density in the Universe, yet the details
of their formation and evolution are still poorly understood. Even in the Milky
Way for which we have access to a tremendous amount of information, the onset of
star formation and the evolution of the star formation rate (SFR) of each
Galactic component are still a matter of debate.
However, the details of the formation of stellar systems are encoded in the
distribution of the stars in deep colour-magnitude diagrams (CMDs). Their star
formation history (SFH), that is, the evolution of both the SFR and the
metallicity from the earliest epoch to the present time, can therefore be
recovered using the robust CMD-fitting technique which has been extensively validated in studies of nearby Local Group galaxies. This technique requires the
precise knowledge of the intrinsic luminosity of each star, and therefore
its distance. In the coming years, \emph{Gaia} will deliver distances and proper
motions for over a billion stars out to $\sim$10 kpc, thus covering all the
structural components of our Galaxy. For the first time, this opens the
possibility of mapping the spatial and temporal variations of the SFH back to
the earliest epochs.
We illustrate this potential by exploiting the Gaia DR1/TGAS parallaxes for stars in the solar neighbourhood \citep{gai16,lin16}, and calculating the SFH of the Milky Way disc within 250~pc of the Sun.
The main advantages of the CMD-fitting technique over other methods of recovering the SFH (e.g.\ chemical enrichment models, colour-function fitting,
age/metallicity census of individual stars, ...) are the fact that determining the age of a population is much more robust than that of single stars, that one takes full advantage of the predictions of stellar evolution models, and the smaller number of assumptions. On the other hand, like other methods it is affected by the systematic effects due to uncertainties in the stellar models, as well as the poorly constrained amplitude of radial migrations in the disc.

\section{The solar neighbourhood CMD: photometry and completeness}
%%----------------------------------------------------------------

While TGAS provides accurate parallaxes and G-band magnitudes for over 2 million stars, no colour information is available. On the other hand, the \emph{Tycho-2} catalogue does include $B_T, V_T$ for all TGAS stars, but the photometric quality quickly degrades at fainter magnitudes. We thus cross-matched TGAS with the \emph{Tycho-2}, \emph{Hipparcos}, and APASS DR9 \citep{hen12} catalogues, after transforming their photometry to the Johnson $B$ and $V$ filters. For stars appearing in more than one catalogue, a weighted mean magnitude was calculated for each filter.

\begin{figure}[t]
 \centering
 \includegraphics[width=4.5in]{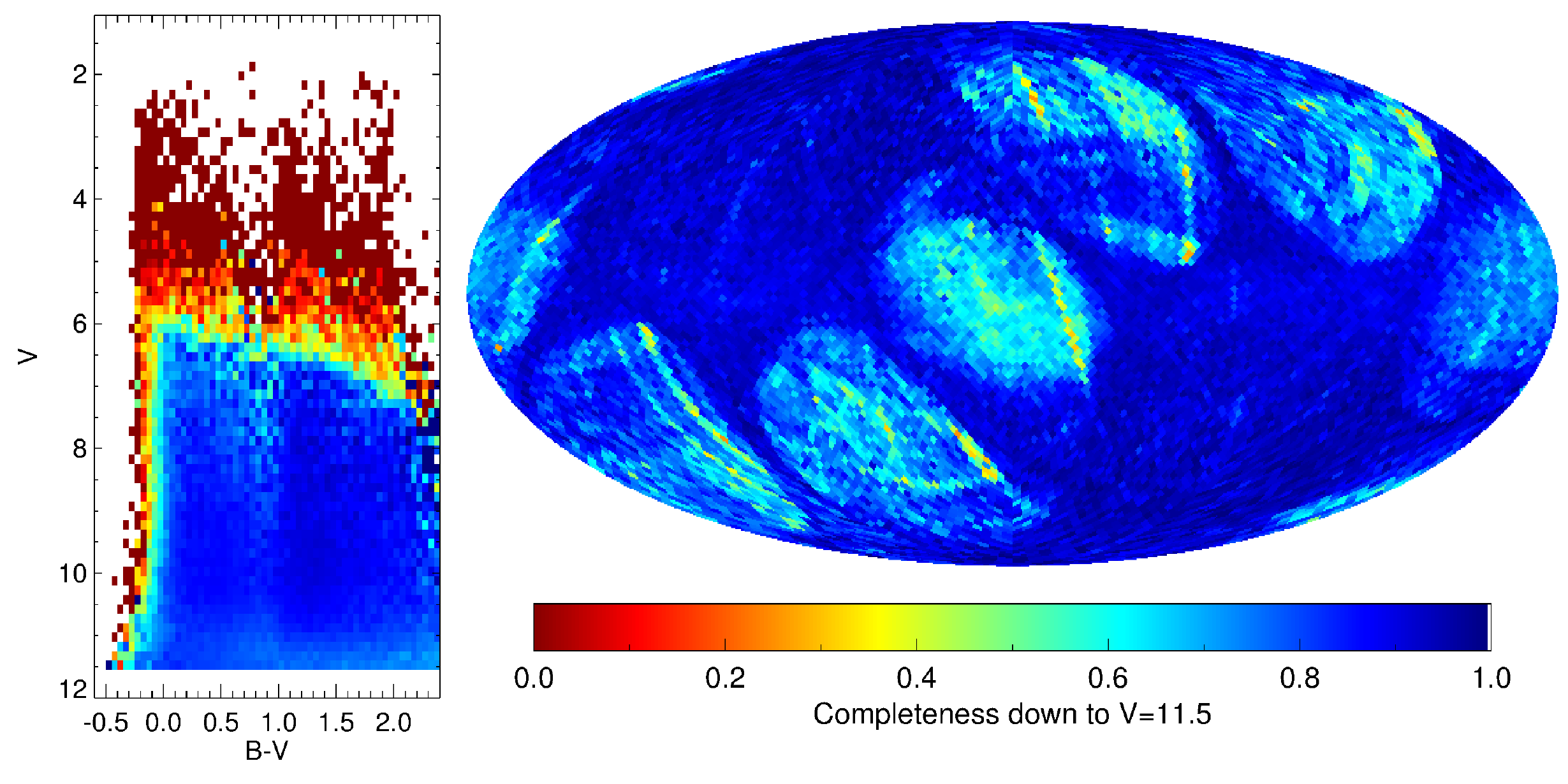}
 \includegraphics[width=2in]{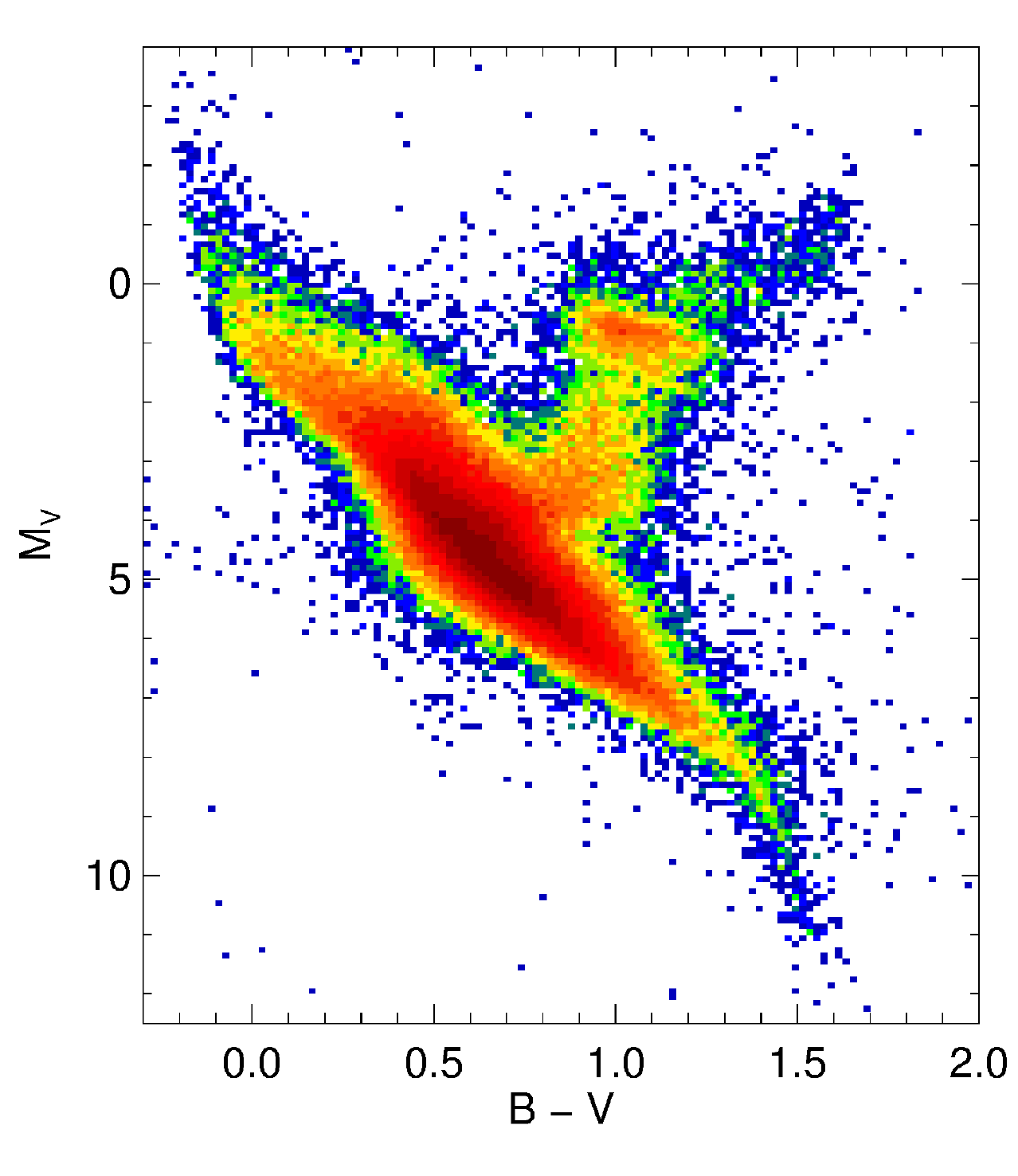}
 \caption{Left \& middle: TGAS completeness down to $V$=11.5 relative to \emph{Tycho-2} in colour-magnitude space and in galactic coordinates, before the completeness corrections. Right: CMD for the solar neighbourhood within 250~pc, which is $\sim$94\% complete down to the magnitude of the oMSTO (M$_V$=4.5).}
   \label{fig1}
\end{figure}

A further step before the CMD-fitting can be applied is a robust quantification of the completeness as a function of colour and magnitude. According to \citet{hog00}, the \emph{Tycho-2} completeness is over 90\% down to $V\sim$\,11.5, but decreases quickly at fainter magnitudes, so we only kept stars brighter than this limit. Even though TGAS is based on \emph{Tycho-2}, about 20\% of the stars from the latter catalogue are missing in TGAS, which means that we have no robust parallaxes for these stars. This is illustrated in Figure\,\ref{fig1}, which presents the completeness of TGAS versus \emph{Tycho-2} as a function of both colour--magnitude (left) and spatial coordinates (middle). The left panel shows that a significant fraction of the stars brighter than $V\sim$\,6.5 or bluer than $B-V$$\sim$0 are missing in TGAS; however, most of these stars have \emph{Hipparcos} parallaxes, which we combined with those from TGAS to erase completeness variations as a function of colour and magnitude.
The middle panel shows that completeness is below $\sim$60\% in about half of the sky due to the lower number of \emph{Gaia} transits. Unfortunately, the completeness function in these regions is too complex to correct; we therefore simply excised 57\% of the sky coverage where completeness down to $V\sim$\,11.5 was $<$90\%.

Finally, to obtain an accurate SFH back to the earliest epoch of star formation, a CMD reaching the oldest main-sequence turn-off (oMSTO, at $M_V$=4.5) is required. Given the completeness limits described above, the volume in which the SFH can be calculated is therefore limited to a distance modulus of ($V-M_V$)=7, corresponding to 250~pc. The resulting CMD, shown in the right panel of Figure\,\ref{fig1}, contains $\sim$148,000 stars and is mostly complete down to the oMSTO within 250~pc.

\section{SFH calculation}
%%-----------------------

The preliminary SFH has been calculated using the technique of synthetic CMD-fitting following the methodology presented in \citet{ber12,ber15a,ber15b}.
The synthetic CMD from which we extracted the simple stellar populations' CMDs
is based on the BaSTI stellar evolution library \citep{pie04}. It contains $2\times10^7$ stars and was generated with a constant SFR over wide ranges of age and metallicity: 0 to 15 Gyr old and
0.0001 $\leq$ Z $\leq$ 0.03 (i.e.\ -2.3 $\leq$ [Fe/H] $\leq$ 0.26, assuming
Z$_{sun}$ = 0.0198; \citealt{gre93}). We adopted a \citet{kro02} initial mass function, and assumed a fraction of unresolved binary systems in TGAS of 10\% with mass ratios between 0 and 1. Further tests with different fractions of binaries, a wider range of metallicity, and different prescriptions for the simulated photometric and parallactic uncertainties are necessary to better understand the possible systematic uncertainties.

While the full photometric uncertainties due to various observational effects are typically estimated using artificial stars tests on the original images \citep[e.g.][]{gal99}, this approach is impossible in the case of \emph{Gaia} for which (most of) the images are not sent back to Earth. Instead, we relied on the distributions of photometric errors as a function of colour and magnitude provided in the \emph{Tycho-2}, \emph{Hipparcos}, and APASS catalogues to simulate the uncertainties in the synthetic CMD.

\begin{figure}[t]
 \centering
 \includegraphics[width=2.47in]{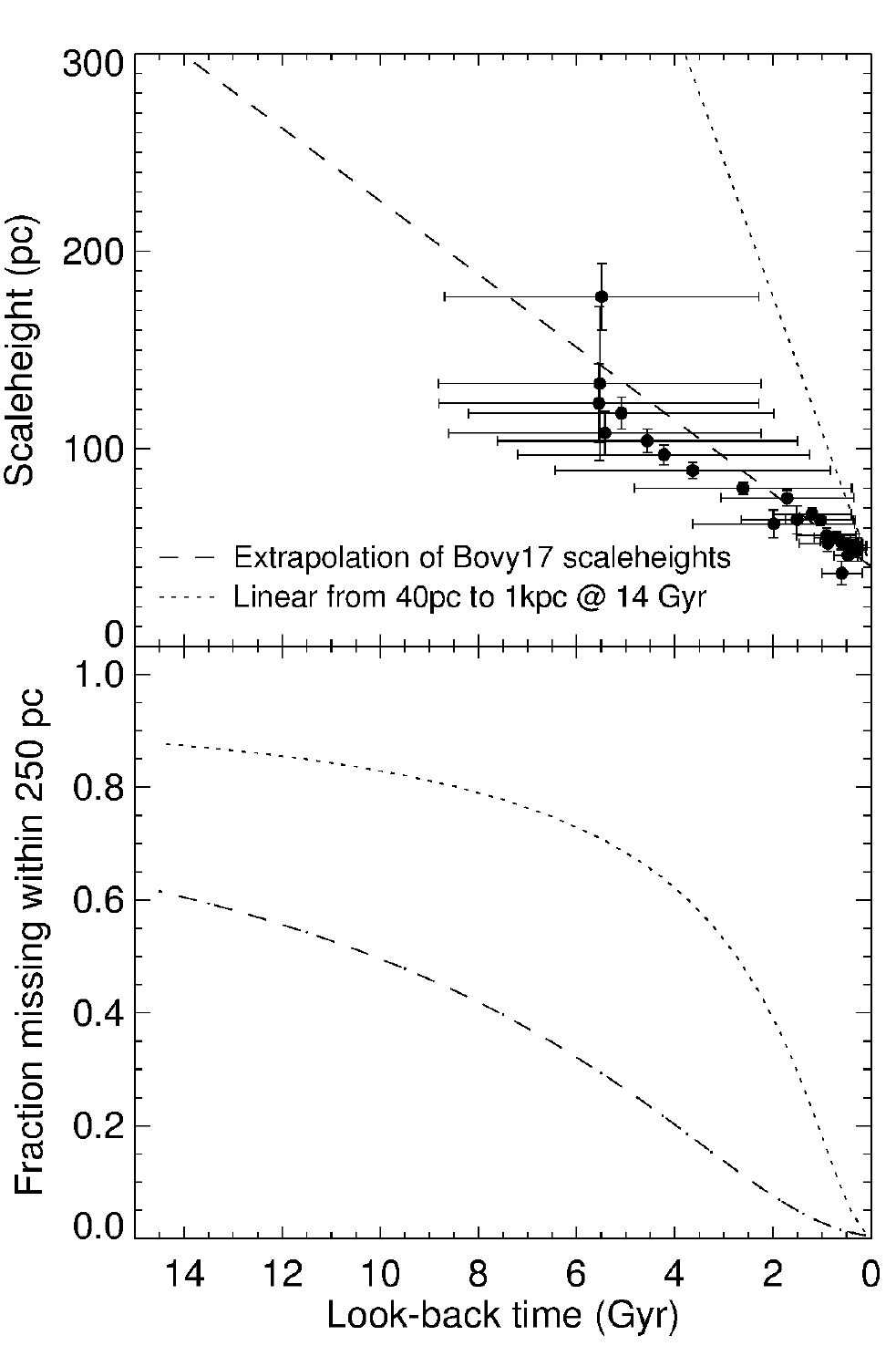}
 \includegraphics[width=2.66in]{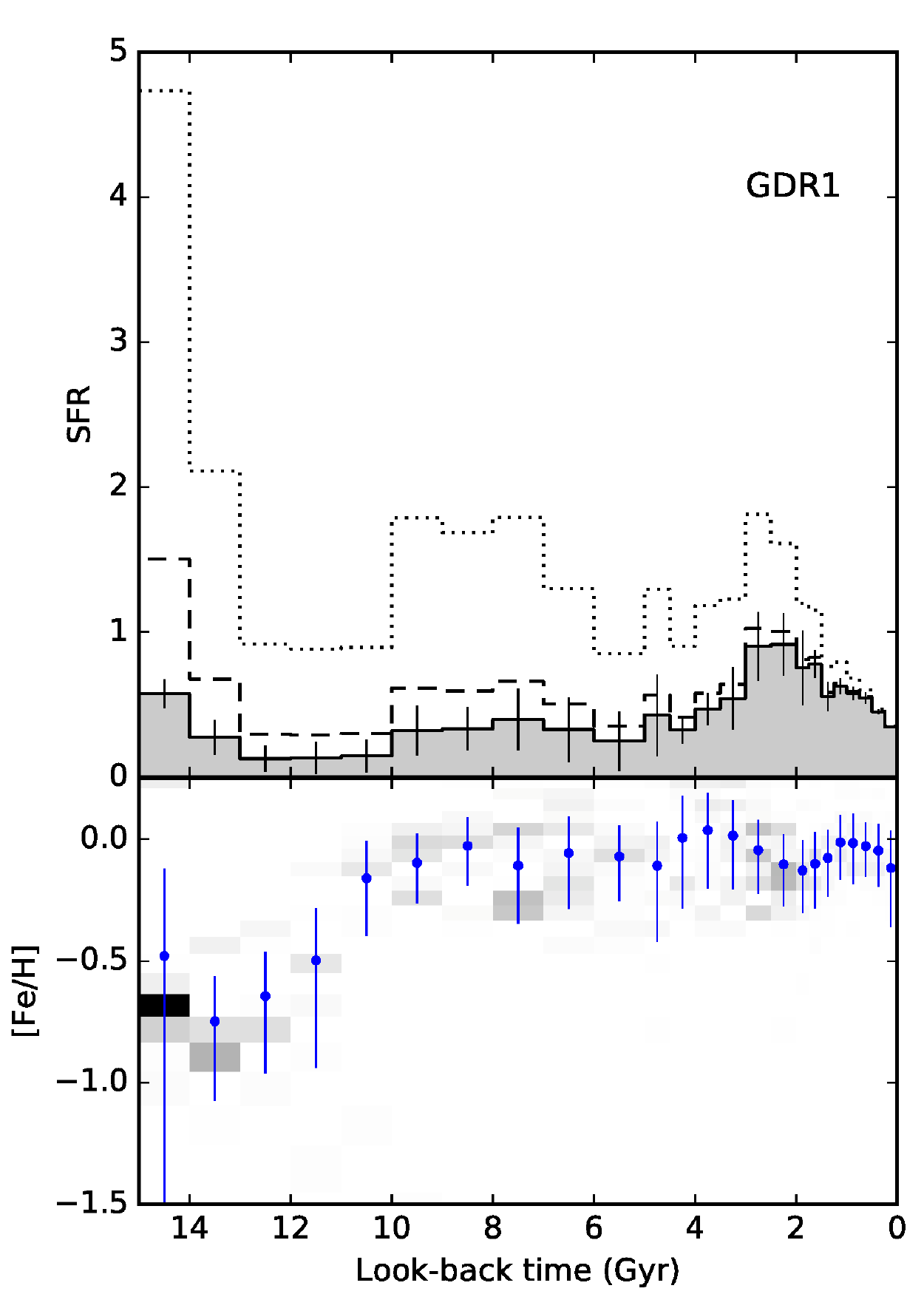}
 \caption{Left: Evolution with time of the disc scaleheight (top) and
 of the corresponding fraction of stars lying beyond 250 pc (bottom, see text
 for details). The dashed and dotted lines correspond to the mild and
 the extreme corrections respectively.
 Right: Resulting SFH, showing the evolution of the SFR (top) and
 metallicity (bottom) as a function of time. In the top panel, the dashed and
 dotted lines represent the SFR corrected for the fraction of stars that have
 been heated to heights $>$250 pc, assuming the mild and the extreme corrections
 respectively.}
   \label{fig2}
\end{figure}

\section{Results}
%%---------------

The SFH is presented in the right panel of Figure\,\ref{fig2}: the top and bottom plots show the evolution of the \emph{dynamically-evolved SFR} (grey histogram) and metallicity, respectively, as a function of time.
However, since the SFH was reconstructed based on the stars that are located \emph{today} within the solar neighbourhood, the effects of secular evolution of the disc have to be taken into account. While the importance of radial migrations (e.g.\ \citealt{sel02}) has yet to be quantified, we can correct this SFR for the disc thickening with age to obtain a SFR that is representative of the whole solar cylinder.

To quantify the thickening, we used the disc scaleheight measured by \citet{bov17} for each spectral type from A0 to G3. The mean age of each type was estimated from the synthetic CMD of the solar neighbourhood from the Besan\c con model \citep{rob03} by selecting all the stars with a temperature within 2\% of the spectral type temperature \citep[from][]{pec13}. The solid circles in the top left panel of Figure\,\ref{fig2} show the \citet{bov17} scaleheights plotted as a function of our estimated ages, where the horizontal errorbars represent the age standard deviation. The dashed line fitted to these points -- virtually the same relation as that used by \citet{jus10} -- shows a clear change of the disc thickness as a function of age, though it could either imply that the older disc formed thicker or that it thickened with time.
We then used the scaleheight fit to estimate the fraction of stars of a given age that are beyond 250 pc and therefore missing from our CMD; this is shown in the bottom left panel of Figure\,\ref{fig2}. The SFR including this correction (hereafter the \emph{mild} correction) is shown as a dashed line in the top right panel.

Note, however, that extrapolating the \citet{bov17} scaleheights to 14 Gyr ago implies an old disc that is only $\sim$300 pc thick, while the Milky Way thick disc is believed to be about 1~kpc thick \citep[e.g.][]{jur08}. Therefore, in the top left panel of Figure\,\ref{fig2} we also show as dotted line a scaleheight increasing linearly from 40 pc at the present day to 1\,kpc 14~Gyr ago, the corresponding fraction of missing stars in the bottom panel, and the resulting SFR in the top right panel.
We label it the \emph{extreme} correction as it clearly over-estimates the scaleheight of the intermediate-age populations; the true relation is likely more complex, possibly with a break around 10 Gyr ago corresponding to the formation of the thick disc.

The dynamically-evolved SFR (i.e.\ uncorrected; grey histogram) shows a constant SFR for the first 10~Gyr or so, with a slight enhancement in the past 4 Gyr. This is in excellent agreement with the SFR calculated from \emph{Hipparcos} data within a smaller volume \citep[$\sim$80~pc;][]{ver02,cig06}. The SFR with the \emph{mild} correction is not significantly different, except perhaps for the slightly more pronounced enhancement at early epochs corresponding to the formation of the thick disc.
On the other hand, with the \emph{extreme} correction the SFR appears roughly constant over most of the history, but with a strong enhancement at early epochs and a decreasing SFR in the past 2--3 Gyr.
This shows that even with the most extreme correction our results are not consistent with an exponentially decreasing SFR as found by several recent studies \citep{aum09,jus11,bov17}. Instead, it favors solutions with a roughly constant star formation over 8--10 Gyr such as found by, e.g., \citet{sna15}.

The age-metallicity relation (AMR), shown in the bottom-right panel of Figure\,\ref{fig2}, is mostly flat for the past 10~Gyr. Only the oldest stars show a lower mean metallicity ([Fe/H]$\sim-$0.7), which may correspond to the thick disc population. This is fully consistent with the independent results from other groups using different methods and datasets \cite[e.g.][]{cas11,hay13,ber14}.

\section{Conclusions and future prospects}
%%----------------------------------------

We have used the \emph{Gaia} DR1 parallaxes to produce a deep CMD that is mostly complete down to the magnitude of the oMSTO within 250~pc from the Sun. We applied the CMD-fitting technique to reconstruct the dynamically-evolved SFH of the local Milky Way disc.
Our results are fully consistent with those obtained previously using the \emph{Hipparcos} data, despite the difficulty of dealing with the complex TGAS completeness function and photometric uncertainties from different catalogues.
We then correct this SFR for the disc thickening with age to obtain a SFR that is representative of the whole solar cylinder, and show that even with an extreme correction our results are not consistent with an exponentially decreasing SFR.
We plan to use the same technique with upcoming \emph{Gaia} data releases. With parallaxes and homogeneous photometry in 3 bands ($G$, $BP$, $RP$) for $>$$10^9$ stars, and not limited by the poorly understood completeness function of an input catalogue like \emph{Tycho-2} was, it will allow us to extend this analysis out to about 5~kpc, and therefore to quantify the SFH of the thin disc, thick disc and halo \emph{in situ}, and its spatial variations.
The spatial variations of the SFH within each component will also provide important constraints on the dynamical processes involved in shapping up their current stellar content.

\begin{acknowledgements}
I would like to thank Lia Athanassoula, James Binney, Cristina Chiappini,
Fran\c coise Combes, Annette Ferguson, Carme Gallart, Andreas Just, and
Annie Robin for useful discussions and comments,
and acknowledge support from the CNES postdoctoral fellowship program.
This work has made use of data from the European Space Agency (ESA)
mission {\it Gaia} (https://www.cosmos.esa.int/gaia), processed by
the {\it Gaia} Data Processing and Analysis Consortium (DPAC,
https://www.cosmos.esa.int/web/gaia/dpac/consortium). Funding for
the DPAC has been provided by national institutions, in particular the
institutions participating in the {\it Gaia} Multilateral Agreement.
\end{acknowledgements}

\end{document}